\documentclass{article}
\usepackage[a4paper,total={16cm,24cm},centering]{geometry}
\usepackage{authblk}
\usepackage{amsmath,amssymb}
\usepackage{graphicx}
\usepackage{subcaption}
\usepackage{braket}
\usepackage{pythonhighlight}
\usepackage{url}
\usepackage{pgfplots}
\pgfplotsset{width=10cm,compat=1.9}
\usepackage{tikz}
\usepackage[acronym]{glossaries}
\usepackage{algorithm,algpseudocode}
\usepackage{svg}
\usepackage[hidelinks]{hyperref}
\usepackage[capitalise]{cleveref}
\usepackage{float}
\usepackage{quantikz}
\usepackage{tikzscale}

\usetikzlibrary{calc}
\definecolor{graphnodecolorA}{HTML}{BECEE1}
\definecolor{graphnodecolorB}{HTML}{B8D697}
\definecolor{graphnodecolorC}{HTML}{FE8585}
\tikzstyle{graphnode}=[circle,font=\small\sffamily,text=black,inner sep=0pt,outer sep=0pt, minimum width=.5cm]
\tikzstyle{graphnodeA}=[graphnode,fill=graphnodecolorA]
\tikzstyle{graphnodeB}=[graphnode,fill=graphnodecolorB]
\tikzstyle{graphnodeC}=[graphnode,fill=graphnodecolorC]
\tikzstyle{graphedge}=[draw=black,thick]
\definecolor{circuithcolor}{HTML}{E55934}
\definecolor{circuitrzcolor}{HTML}{5BC0EB}
\definecolor{circuitrxcolor}{HTML}{502C65}
\definecolor{circuitcxcolor}{HTML}{075678}
\tikzstyle{circuitline}=[draw=black]
\tikzstyle{circuitq}=[draw=none,font=\small\sffamily]
\tikzstyle{circuith}=[fill=circuithcolor,inner sep=0pt,outer sep=0pt,text=black,font=\footnotesize\sffamily,minimum width=.45cm,minimum height=.45cm]
\tikzstyle{circuitrz}=[fill=circuitrzcolor,inner sep=0pt,outer sep=0pt,text=black,font=\scriptsize\sffamily,minimum width=.45cm,minimum height=.45cm]
\tikzstyle{circuitrx}=[fill=circuitrxcolor,inner sep=0pt,outer sep=0pt,text=white,font=\scriptsize\sffamily,minimum width=.45cm,minimum height=.45cm]
\tikzstyle{circuitcx1}=[circle,fill=circuitcxcolor,inner sep=0pt,outer sep=0pt,minimum width=.15cm]
\tikzstyle{circuitcx2}=[circle,fill=circuitcxcolor,inner sep=0pt,outer sep=0pt,text=white,font=\footnotesize\sffamily]
\tikzstyle{circuitcxl}=[thick,draw=circuitcxcolor]
\newcommand{\circuith}[2]{%
\node[circuith] (h) at ($(#1)+(.15,0)+#2*(.55,0)$) {$H$};%
}
\newcommand{\circuitrz}[2]{%
\node[circuitrz] (rz) at ($(#1)+(.15,0)+#2*(.55,0)$) {$R_Z$};%
}
\newcommand{\circuitrx}[2]{%
\node[circuitrx] (rx) at ($(#1)+(.15,0)+#2*(.55,0)$) {$R_X$};%
}
\newcommand{\circuitcx}[3]{%
\node[circuitcx1] (cx1) at ($(#1)+(.15,0)+#3*(.55,0)$) {};%
\node[circuitcx2] (cx2) at ($(#2)+(.15,0)+#3*(.55,0)$) {+};%
\draw[circuitcxl] (cx1) -- (cx2);%
}
\allowdisplaybreaks

\newacronym{dqc}{DQC}{Distributed Quantum Computing}
\newacronym{mbqc}{MBQC}{Measurement-Based Quantum Computing}
\newacronym[plural=QPUs,firstplural=Quantum Processing Units]{qpu}{QPU}{Quantum Processing Unit}
\newacronym{nisq}{NISQ}{Noisy Intermediate-Scale Quantum}
\newacronym{qaoa}{QAOA}{Quantum Approximate Optimization Algorithm }

\title{Bipartitioning of Graph States for Distributed Measurement-Based Quantum Computing}
\author[1]{Kjell Fredrik Pettersen}
\author[2,3]{Matthias Heller}
\author[1]{Giorgio Sartor\thanks{Corresponding author: \href{mailto:giorgio.sartor@sintef.no}{giorgio.sartor@sintef.no}}}
\author[4]{Raoul Heese}
\affil[1]{SINTEF Digital, Oslo, Norway}
\affil[2]{Fraunhofer Institute for Computer Graphics Research IGD, Darmstadt, Germany}
\affil[3]{Technical University of Darmstadt, Interactive Graphics Systems Group, Darmstadt, Germany}
\affil[4]{NTT DATA, Munich, Germany}

\date{\today}

\begin{document}

\maketitle

\begin{abstract}
    \Gls{mbqc} is inherently well-suited for \gls{dqc}: once a resource state is prepared and distributed across a network of quantum nodes, computation proceeds through local measurements coordinated by classical communication. However, since non-local gates acting on different \glspl{qpu} are a bottleneck, it is crucial to optimize the qubit assignment to minimize inter-node entanglement of the shared resource. For graph state resources shared across two \glspl{qpu}, this task reduces to finding bipartitions with minimal cut rank. We introduce a simulated annealing-based algorithm that efficiently updates the cut rank when two vertices swap sides across a bipartition, such that computing the new cut rank from scratch, which would be much more expensive, is not necessary. We show that the approach is highly effective for determining qubit assignments in distributed \gls{mbqc} by testing it on grid graphs and the measurement-based \gls{qaoa}.
\end{abstract}
\glsresetall

\section{Introduction}
Current quantum computing hardware platforms face significant challenges in scaling up the number of qubits on a single \gls{qpu}. 
As the qubit count increases, issues such as decoherence, gate fidelity, control complexity, and physical infrastructure limitations make it increasingly difficult to maintain performance and reliability.
Unfortunately, these constraints limit the practical use of quantum computing, particularly when fault tolerance is required.
\Gls{dqc} has been recently presented as a viable option to scale up the number of available qubits in the future, for recent reviews see, e.g.,~\cite{CALEFFI2024110672,boschero2025distributed,BARRAL2025100747}. 
The basic idea of \gls{dqc} is to link together several \glspl{qpu} giving rise to a quantum network.
However, non-local gates implemented through such quantum links typically have significantly lower fidelity as well as longer execution time compared to the on-chip operations~\cite{nickerson2013,gupta2025}.
The usual assumption for any \gls{dqc} protocol is that there is a limited number of  Bell states (EPR pairs) shared between different \glspl{qpu}, which are used (consumed) to implement quantum operations between the nodes either by qubit or gate teleportation~\cite{BARRAL2025100747}. 
The amount of these shared pairs is limited.
Therefore, a new challenge arises in the setting of \acrfull{dqc}: how to compile a quantum circuit such that the number of EPR pairs needed to implement an algorithm is minimized?

\Gls{mbqc} is a an alternative model of quantum computing that consists of preparing an initial, highly-entangled quantum state (called resource state) and then drive its evolution through consecutive, adaptive single-qubit measurements~\cite{raussendorf2001one}. 
What makes \gls{mbqc} particularly interesting in the context of \gls{dqc} is that the initial resource state can be distributed at the start of the computation, such that the expensive non-local gates can be performed right in the beginning. 
After that, the protocol only uses classical communication to report the outcome of local measurements on the individual \glspl{qpu} to implement the round of adaptive measurements.
Distributed \gls{mbqc} has been discussed and formalized in the literature~\cite{DANOS200773}.

Typically, the resource state chosen to perform \gls{mbqc} is a graph state, which---as its name suggests---can be associated with an undirected graph in which vertices represent qubits and edges represent entanglement between pairs of qubits. 
Specifically, the graph state associated with an undirected graph $G=(V,E)$ with vertices $V$ and edges $E$ reads
\begin{equation} \label{eqn:G}
    \ket{G}:=\prod_{e \in E} CZ_e \ket{+}^{\otimes |V|},
\end{equation}
where $CZ_e$ represents a controlled-Z gate on two qubits connected by edge $e$, and $\ket{+}$ is the eigenstate of the Pauli-X operator with eigenvalue +1.
Since controlled-Z gates are symmetric and commute, neither the qubit order in $e$ nor the product order in \cref{eqn:G} matter.
The application of local Clifford gates to $\ket{G}$ can be understood as a transformation of the associated graph, where the changes to the graph can be formalized as a sequence of local complementations (and vertex relabelings)~\cite{van2004graphical,PhysRevA.73.022334}. 
To be more precise, the local complementation $G*v$ of a graph $G$ at a vertex $v \in V$ is defined by complementing the neighborhood $N(v)$ of $G$, which means that we remove an edge $(i,j)\in E$ from $G$ if $i,j\in N(v)$ or add the edge, if it wasn't there before.
The result of a local complementation $G'=G*v$ is a state~\cite{PhysRevA.73.022334}
\begin{equation}
    \ket{G'} = \sqrt{-iX_v} \prod_{k \in N(v)}\sqrt{i Z_k} \ket{G},
\end{equation}
which is locally equivalent to $\ket{G}$ since local gates do not change the bipartite entanglement.
Consequently, the corresponding graphs can also be considered locally equivalent, meaning they can be reconfigured via local complementations while preserving computational equivalence of the \gls{mbqc} procedure on the respective graph states.

From the \gls{mbqc} viewpoint, distributing a computation across multiple \glspl{qpu} reduces to a graph partitioning problem:
choose a partition of $V$ such that each part fits on a \gls{qpu} and the \emph{cut rank}, which equals the Schmidt rank across the partition for graph states~\cite{hein2004multiparty,Nest:2006rge}, is minimized.
Here and in the following, we limit ourselves to a \gls{dqc} scenario with exactly two \glspl{qpu}.
The cut rank of a graph $G=(V,E)$, given the bipartition into $X\subseteq V$ and $Y=V\setminus X$, is defined as~\cite{nguyen2020}
\begin{equation}
    {\rm \rho}_{X,Y}(G):={\rm rank}\big(A[X,Y]\big),
\end{equation}
where $A[X,Y]$ denotes the adjacency matrix of $G$ with only the rows $X$ and columns $Y$ defined over the binary field $GF(2)$.
In short, finding a partition $(X,Y)$ of minimal cut rank corresponds to an optimal partition of the qubits in the resource state.

The cut rank of a bipartition measures how many rows (or columns) are linearly independent, so it captures the diversity of connections, not just the quantity.
Indeed, in the case when the amount of edges connecting two partitions is larger than the cut rank, one can reduce the number by embedding the graph into a slightly larger one, which is equivalent up to local complementations and vertex deletions, which in the \gls{mbqc} picture correspond to measurements of qubits. 
Since the bipartite Schmidt rank is invariant under local unitaries (in particular local Cliffords), the cut rank is invariant under local complementations~\cite{dum2017}.

In this paper, we develop a heuristic algorithm to find the bipartition of a graph with the smallest cut rank using a simulated annealing approach.
Given a graph $G=(V,E)$ and a bipartition $(X,Y)$ with $X\subseteq V$ and $Y=V\setminus X$, we develop an algorithm that computes the change in cut rank for a fixed vertex for every possible swap with vertices from the other set in $O(n^2)$. 
We use this algorithm to accelerate simulated annealing-based optimization~\cite{Kirkpatrick}, for which computing the change in cut rank would be the computational bottleneck.
Here, we propose a simple analytical and efficient method that cuts down the computation time by up to two-orders of magnitude for graphs with up to $400$ nodes.

In \cref{sec:distMBQC}, we provide a brief summary of \gls{mbqc} for \gls{dqc} that motivates the search for optimal partitions. Subsequently, we present the heuristic algorithm in \cref{sec:algorithm}, which is the main contribution of this work. In \cref{sec:experiments}, we verify the algorithm with numerical experiments. Finally, we close with conclusions and and outlook in \cref{sec:conclusions}.

\section{Distributing graph states for MBQC}
\label{sec:distMBQC}

In this section, we outline how finding a partition of a graph with small cut rank can be used to cut the graph state used as resource state in \gls{mbqc} into smaller pieces such that a larger computations can be distributed with a minimal number of shared Bell states.
Specifically, we assume, that we want to implement a \gls{mbqc} computation in standard form~\cite{broadbent2009parallelizing}, which consists of three stages:
\begin{enumerate}
    \item Graph state preparation,
    \item adaptive measurements of ancillary qubits, and
    \item a final round of Pauli corrections on the output qubits.
\end{enumerate}
For more details on how this form can be achieved, we refer to, e.g., \cite{kaldenbach2023,Kaldenbach:2025jle,Vijayan2024Compilation}.
The partition strategy presented in this paper only concerns the first stage, as this is the part in which entanglement is shared between the different \glspl{qpu}. If this first step is achieved, the other two stages can be performed straightforwardly and require only classical communication between the \glspl{qpu}.

\paragraph{Distribution rule}
Let $G(V,E)$ be the graph corresponding to the resource state on which we perform our measurements and let $(X,Y)$ with $X \in V$ and $Y=V\setminus X$ be a bipartition with cut rank $r$.
Then we can distribute the state across two \glspl{qpu} requiring $r$ EPR pairs shared between them. 

\paragraph{Proof}
The distribution rule is based on Lemma~3.3 of \cite{campbell2025erdhosposapropertycirclegraphs}, which we sketch here. 
If the cut rank of the bipartition $(X,Y)$ is $r$, we know that the adjacency matrix $A[X,Y]$ has rank $r$ and therefore can be written as a sum over $r$ rank-one matrices:
\begin{equation}
    A[X,Y] = \sum_{i=1}^r A_i[X_i,Y_i],
\end{equation}
where $X_i$ and $Y_i$ are subsets of $X$ and $Y$.
For each term $A_i$, we introduce two ancillary qubits, $q_i^a$ and $q_i^b$, of which we connect one with all $i\in X_i$ and the other with all $i\in Y_i$.
If we now perform a local complementation over all $q_i^a$, another over $q_i^b$, and a third over $q_i^a$, we get back the original graph after deletion of all ancillary qubits.
This operation corresponds to a measurement of $q_i^a$ and $q_i^b$ in the $X$ basis~\cite{PhysRevA.73.022334,hein2004multiparty}.
A similar operation has been introduced in \cite{Hoyer2006Resources}, where it was used to reduce the degree of a given graph by embedding it into a larger one.

\begin{figure}[htb]
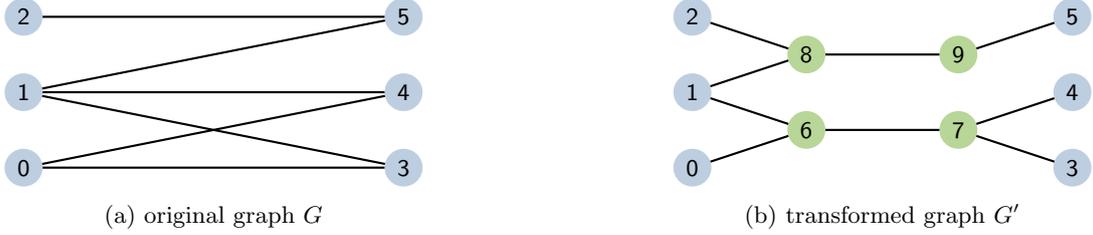

    \centering
    \begin{subfigure}[t]{.45\textwidth}
    \centering
    \includegraphics{graph_example.tikz}
    \caption{original graph $G$}
    \label{fig:graph_embedding:G1}
    \end{subfigure}
    \hfill
    \begin{subfigure}[t]{.45\textwidth}
    \centering
    \includegraphics{graph_example_embedded.tikz}
    \caption{transformed graph $G'$}
    \label{fig:graph_embedding:G2}
    \end{subfigure}
    \caption{Example of how to embed a partitioned graph such that the cut rank of its partition corresponds to the number of edges between them. The transformation between the original graph $G$ and the transformed graph $G'$ can be achieved with local transformations and node removals as defined in \cref{eqn:G:example}. In the context of \gls{dqc}, both graphs are locally equivalent and demonstrate how the entanglement connections are shared across two \glspl{qpu}. For $G'$, the qubits $6$ to $9$ correspond to two EPR pairs.}
    \label{fig:graph_embedding}
\end{figure}

\paragraph{Example}
Let $G(V,E)$ be a graph with vertices $V:=\{0,\dots,5\}$ and edges 
\begin{equation}
E:=\{(0,3),(0,4),(1,3),(1,4),(1,5),(2,5)\}.
\end{equation}
The bipartition into $X=\{0,1,2\}$ and $Y=\{3,4,5\}$ has cut rank $2$.
We introduce four ancillary nodes $\{6,7,8,9\}$, and define $G'\left(V\cup\{6,7,8,9\}, E'\right)$, where
\begin{equation}
E':=\{(0,6),(1,6),(6,7),(7,3),(7,4),(2,8),(1,8),(8,9),(9,5)\}.
\end{equation}
Local complementation sequences over the nodes $6$, $7$, $6$ and $8$, $9$, $8$ as well as the final removal of the ancillary nodes $\{6,7,8,9\}$ reproduces the original graph. That is,
\begin{equation} \label{eqn:G:example}
G=G'*6*7*6*8*9*8 - \{6,7,8,9\}.
\end{equation}
This example is shown in \cref{fig:graph_embedding}.

\section{An incremental algorithm for cut rank}
\label{sec:algorithm}

As explained in the previous section, the cut rank of a graph bipartition directly relates to the number of shared Bell states needed to implement distributed \gls{mbqc}. Hence, finding a partitioning with low cut rank is key for an efficient realization of \gls{dqc}.
The aim of the present section is to develop an algorithm that takes a simple graph $G=(E,V)$ and partition size $n$ as input and finds a bipartition $X\subseteq V$ and $Y=V\setminus X$ with minimal cut rank and such that $|X|=n$.
The basic idea is to use simulated annealing~\cite{Kirkpatrick} to solve this problem as defined in \cref{alg:sa}.

\begin{algorithm}[H] 
\caption{Simulated annealing for fixed-size partitioning} \label{alg:sa}
\begin{algorithmic}[1]
\Require Graph $G=(V,E)$, partition size $n$, temperature schedule $\mathcal{T}=T_1,T_2,\dotsc T_N$
\State \textbf{Initialize:} Select a random initial partition $X$ with $|X|=n$.
\State $c={\rm rank}(A[X,V\setminus X])$
\For{$T_i \in \mathcal{T}$}
\For{$i \in X$}
\For{$j \in V\setminus X$}
        \State $\tilde X = \big(X\setminus \{i\} \big)\cup \{j\}$
        \State $\Delta c = {\rm rank}(A[\tilde X,V\setminus \tilde X]) - c$
        \If{$\exp(-\Delta c/T_i) > {\rm rand}(0,1)$}
        \State $X\gets \tilde X$
        \State $c \gets c+\Delta c$
        \EndIf
\EndFor
\EndFor
\EndFor
\end{algorithmic}
\end{algorithm}

The most computationally demanding step in this algorithm is in line $7$, where we have to calculate the change of cut rank after performing one swap of vertices in the two partitions.
Naively, one would first calculate the new cut rank corresponding to $\tilde X$ and then calculate $\Delta c$, which would require complexity $O(n^3)$ due to the rank calculation.
However, $\Delta c$ can be calculated more efficiently by using information from the previous calculations.
This can be done by storing some key matrices defined from the current partition from which one can calculate the cut rank changes for a fixed $i$ simultaneously for all $j$ with time complexity $O(n^2)$. 
The complexity of maintaining the necessary key matrices after applying a swap also has complexity $O(n^2)$.

\subsection{Key matrices in the cut rank calculations}

The cut rank for the current partition is given as ${\rm \rho}_{X,Y}(G)={\rm rank}(A[X,Y])$.
This implies there are subsets $X^B\subseteq X$ and $Y^B\subseteq Y$ where $|X^B|=|Y^B|={\rm \rho}_{X,Y}(G)$ such that the submatrix
\begin{equation}
    C := A[X^B,Y^B]
\end{equation}
of $A[X,Y]$ is invertible. The sets $X^B$ and $Y^B$ are assumed to be known, as well as the inverse $C^{-1}$ which is naturally indexed with $Y^B$ as rows and $X^B$ as columns. The rows (resp. columns) indexed by $X^B$ ($Y^B$) define a basis for the row (column) vectors of $A[X,Y]$, therefore
\begin{equation}
    \label{eq:basis_generates_free}
    A[X^F,Y^B]\cdot C^{-1}\cdot A[X^B,Y^F] = A[X^F,Y^F],
\end{equation}
where $X^F=X\setminus X^B$ and $Y^F=Y\setminus Y^B$ are the remaining vertices in the partition sets.

To perform the cut rank calculation we also need the matrices
\begin{equation}
    D_X := A[V(G),Y^B]\cdot C^{-1},
\end{equation}
\begin{equation}
    D_Y := C^{-1}\cdot A[X^B,V(G)],
\end{equation}
and
\begin{equation}
    F := A[V(G),Y^B]\cdot C^{-1}\cdot A[X^B,V(G)] + A[V(G),V(G)],
\end{equation}
where the columns of $D_X$ are indexed by $X^B$ and the rows of $D_Y$ are indexed by $Y^B$.

\subsection{The cut rank calculation process}

A swap that updates the partition sets $X$ and $Y$ by swapping the vertices $i\in X$ and $j\in Y$ may have an effect on the cut rank and basis sets $X^B$ and $Y^B$. The process for finding these updates has two steps, reduction and extension.

\begin{itemize}
\item{\bf Reduction:} The basis sets $X^B$ and $Y^B$ are reduced by equally many vertices so $i$ and $j$ are no longer in the basis sets, and the reduced $A[X^B,Y^B]$ matrix is still invertible. This is only necessary if $i\in X^B$ or $j\in Y^B$. Maximum 2 vertices will be removed from each set.
\item{\bf Extension:} The vertices are swapped, and the new $X^B$ and $Y^B$ sets are extended if necessary with equally many vertices until \cref{eq:basis_generates_free} is satisfied. The sizes of $X^B$ and $Y^B$ give the new cut rank after the swap is applied. Maximum 4 vertices will be added to each basis set, and the net change of the cut rank from the reduction and extension combined will be limited by $\pm 2$.
\end{itemize}

The vertices to be removed and added in the reduction and extension steps are found by matrix investigations and attempts to clean the
matrix $A[X,Y]$ with the basis rows and columns. The process will depend on wether certain properties are satisfied. 
Some of them only depend on one of the swapped vertices, and may be settled for all $i$ and $j$ in a preprocessing phase. Two of these properties are
\begin{align}
    P^X_1(i) &:= (i\in X^B)\wedge(\exists k_1\in X^F:D_X[k_1,i]=1)\nonumber\\
    P^Y_1(j) &:= (j\in X^B)\wedge(\exists l_1\in Y^F:D_Y[l_1,j]=1) 
\end{align}
If $P^X_1(i)$ (resp. $P^Y_1(j)$) is true we assume $k_1$ ($l_1$) is implicitly given.

The formulation of the next properties depend on the truth values of the first ones:
\begin{equation}
    P^X_2(i) := \exists k_2\in X^F:
    \begin{cases}
        \begin{aligned}
            & (k_2\neq k_1)\wedge(F[k_2,i] \\
            & \quad\quad\quad +F[k_1,i]D_X[k_2,i]=1)
        \end{aligned} & \text{if $P^X_1(i)$} \\
        (k_2\neq i)\wedge(F[k_2,i]=1) & \text{if not $P^X_1(i)$}
    \end{cases}
\end{equation}
\begin{equation}
    P^Y_2(j) := \exists l_2\in Y^F:
    \begin{cases}
        \begin{aligned}
            & (l_2\neq l_1)\wedge(F[j,l_2] \\
            & \quad\quad\quad +F[j,l_1]D_Y[j,l_2]=1)
        \end{aligned} & \text{if $P^Y_1(j)$} \\
        (l_2\neq j)\wedge(F[j,l_2]=1) & \text{if not $P^Y_1(j)$}
    \end{cases}
\end{equation}
Again, if $P^X_2(i)$ (resp. $P^Y_2(j)$) is true we take $k_2$ ($l_2$) implicitly as given.

The matrix $C^{-1}$ is invertible. It has therefore no row or column with $0$ only. This implies
\begin{equation}
    i\in X^B \Rightarrow \exists\alpha\in Y^B:C^{-1}[\alpha,i]=1
\end{equation}
\begin{equation}
    j\in Y^B \Rightarrow \exists\beta\in X^B:C^{-1}[j,\beta]=1
\end{equation}
If $i\in X^B$ (resp. $j\in Y^B$), we take $\alpha$ ($\beta$) as given. 
None of the vertices $k_2$, $l_2$, $\alpha$ and $\beta$ are needed to only determine the value of the cut rank, but may occur among the removed or added vertices.

The process is split into five different cases, depending on whether $i$ and $j$ are in the basis sets, and for the case when both are, the value of $C^{-1}[j,i]$. 
The vertices to remove in the reduction step are the same within each of these cases, while there will be several sub-cases for the extension step.

\paragraph{First case: $i\in X^F$ and $j\in Y^F$}
No reduction is needed since the $C$ matrix does not contain row $i$ or column $j$. The different cases for the extension step depend on the properties $P^X_2(i)$ and $P^Y_2(j)$, and the value of $F[j,i]$, this is summarized by
\begin{center}
\begin{tabular}{ c c c | c | c | c}
$P^X_2(i)$ & $P^Y_2(j)$ & $F[j,i]$ & $\Delta c$ & $+X^B$ & $+Y^B$ \\
\hline
T & T & & $+2$ & $j,k_2$ & $i,l_2$ \\
T & F & & $+1$ & $k_2$ & $i$ \\
F & T & & $+1$ & $j$ & $l_2$ \\
F & F & $1$ & $+1$ & $j$ & $i$ \\
F & F & $0$ & $0$ & &
\end{tabular}
\end{center}
where T means true, F means false, $\Delta c$ is the combined cut rank change from the reduction and extension, and $+X^B$ (resp. $+Y^B$) are the vertices added to $X^B$ ($Y^B$).

\paragraph{Second case: $i\in X^B$ and $j\in Y^F$}
The reduction step will need to remove $i$ from $X^B$ and some column from $Y^B$. If $C'$ is the submatrix of $C$ with row $i$ and column $\alpha$ removed, then
\begin{equation}
    \det(C')=\operatorname{Adj}(C)[\alpha,i]=C^{-1}[\alpha,i]=1
\end{equation}
so $C'$ is invertible. The reduction step can therefore be set to removes $i$ from $X^B$ and $\alpha$ from $Y^B$. 
The extension step can be split into two cases depending on the truth value of $P^X_1(i)$:

\subparagraph{$P^X_1(i)$ is true}
\begin{center}
\begin{tabular}{ c c c | c | c | c}
$P^X_2(i)$ & $P^Y_2(j)$ & $Q$ & $\Delta c$ & $+X^B$ & $+Y^B$ \\
\hline
T & T & & $+2$ & $j,k_1,k_2$ & $i,\alpha,l_2$ \\
T & F & & $+1$ & $k_1,k_2$ & $i,\alpha$ \\
F & T & & $+1$ & $j,k_1$ & $\alpha,l_2$ \\
F & F & $1$ & $+1$ & $j,k_1$ & $i,\alpha$ \\
F & F & $0$ & $0$ & $k_1$ & $\alpha$ \\
\end{tabular}
\end{center}
where $Q:=F[j,i]+D_X[j,i]F[k_1,i]$.

\subparagraph{$P^X_1(i)$ is false}
\begin{center}
\begin{tabular}{ c c c c | c | c | c}
$D_X[j,i]$ & $P^X_2(i)$ & $P^Y_2(j)$ & $F[j,i]$ & $\Delta c$ & $+X^B$ & $+Y^B$ \\
\hline
$1$ & T & & & $+1$ & $j,k_2$ & $i,\alpha$ \\
$1$ & F & & & $0$ & $j$ & $\alpha$ \\
$0$ & T & T & & $+1$ & $j,k_2$ & $i,l_2$ \\
$0$ & T & F & & $0$ & $k_2$ & $i$ \\
$0$ & F & T & & $0$ & $j$ & $l_2$ \\
$0$ & F & F & $1$ & $0$ & $j$ & $i$ \\
$0$ & F & F & $0$ & $-1$ & &
\end{tabular}
\end{center}

\paragraph{Third case: $i\in X^F$ and $j\in Y^B$}
This is symmetric to the previous case. The reduction step will remove $\beta$ from $X^B$ and $j$ from $Y^B$. 
The extension step depends on $P^Y_1(j)$:

\subparagraph{$P^Y_1(j)$ is true}
\begin{center}
\begin{tabular}{ c c c | c | c | c}
$P^Y_2(j)$ & $P^X_2(i)$ & $Q$ & $\Delta c$ & $+X^B$ & $+Y^B$ \\
\hline
T & T & & $+2$ & $j,\beta,k_2$ & $i,l_1,l_2$ \\
T & F & & $+1$ & $j,\beta$ & $l_1,l_2$ \\
F & T & & $+1$ & $\beta,k_2$ & $i,l_1$ \\
F & F & $1$ & $+1$ & $j,\beta$ & $i,l_1$ \\
F & F & $0$ & $0$ & $\beta$ & $l_1$ \\
\end{tabular}
\end{center}
where $Q:=F[j,i]+D_Y[j,i]F[j,l_1]$.

\subparagraph{$P^Y_1(j)$ is false}
\begin{center}
\begin{tabular}{ c c c c | c | c | c}
$D_Y[j,i]$ & $P^Y_2(j)$ & $P^X_2(i)$ & $F[j,i]$ & $\Delta c$ & $+X^B$ & $+Y^B$ \\
\hline
$1$ & T & & & $+1$ & $j,\beta$ & $i,l_2$ \\
$1$ & F & & & $0$ & $\beta$ & $i$ \\
$0$ & T & T & & $+1$ & $j,k_2$ & $i,l_2$ \\
$0$ & T & F & & $0$ & $j$ & $l_2$ \\
$0$ & F & T & & $0$ & $k_2$ & $i$ \\
$0$ & F & F & $1$ & $0$ & $j$ & $i$ \\
$0$ & F & F & $0$ & $-1$ & &
\end{tabular}
\end{center}

\paragraph{Fourth case: $i\in X^B$, $j\in Y^B$ and $C^{-1}[j,i]=1$}
As before, $C$ with row $i$ and column $j$ removed is invertible since $C^{-1}[j,i]=1$. Therefore, the reduction step will remove $i$ from $X^B$ and $j$ from $Y^B$. The extension step splits into two cases:

\subparagraph{$P^X_1(i)$ and $P^Y_1(j)$ are true}
\begin{center}
\begin{tabular}{ c c c | c | c | c}
$P^X_2(i)$ & $P^Y_2(j)$ & $Q$ & $\Delta c$ & $+X^B$ & $+Y^B$ \\
\hline
T & T & & $+2$ & $j,k_1,k_2$ & $i,l_1,l_2$ \\
T & F & & $+1$ & $k_1,k_2$ & $i,l_1$ \\
F & T & & $+1$ & $j,k_1$ & $l_1,l_2$ \\
F & F & $1$ & $+1$ & $j,k_1$ & $i,l_1$ \\
F & F & $0$ & $0$ & $k_1$ & $l_1$
\end{tabular}
\end{center}
where $Q:=F[j,i]+D_X[j,i]F[k_1,i]+D_Y[j,i]F[j,l_1]+F[k_1,i]F[j,l_1]$.

\subparagraph{$P^X_1(i)$ or $P^Y_1(j)$ is false}
Two more properties are needed
\begin{align}
    P^X_3(i,j) &= \exists k_3\in X^F:F[k_3,i]+D_X[k_3,i]D_Y[j,i]=1\nonumber\\
    P^Y_3(i,j) &= \exists l_3\in Y^F:F[j,l_3]+D_X[j,i]D_Y[j,l_3]=1
\end{align}
where we take $k_3$ (resp. $l_3$) implicitly for given if $P^X_3(i,j)$ ($P^Y_3(i,j)$) is true. The extension step splits into the following combinations:
\begin{center}
\begin{tabular}{ c c c | c | c | c}
$P^X_3(i,j)$ & $P^Y_3(i,j)$ & $Q$ & $\Delta c$ & $+X^B$ & $+Y^B$ \\
\hline
T & T & & $+1$ & $j,k_3$ & $i,l_3$ \\
T & F & & $0$ & $k_3$ & $i$ \\
F & T & & $0$ & $j$ & $l_3$ \\
F & F & $1$ & $0$ & $j$ & $i$ \\
F & F & $0$ & $-1$ & &
\end{tabular}
\end{center}
where $Q:=F[j,i]+D_X[j,i]D_Y[j,i]$.

\paragraph{Fifth case: $i\in X^B$, $j\in Y^B$ and $C^{-1}[j,i]=0$}
Removing $i$ and $j$ from the basis will give a singular submatrix of $C$ since $C^{-1}[j,i]=0$, so this is not enough for the reduction step. However the submatrix
\begin{equation}
    C^{-1}[\{j,\alpha\},\{i,\beta\}]:=\begin{bmatrix} 0 & 1 \\ 1 & C^{-1}[\alpha,\beta]\end{bmatrix}
\end{equation}
is invertible, then so is the submatrix of $C$ with the same vertices removed from the basis. Therefore the reduction step will remove $i$, $\beta$ from $X^B$ and $j$, $\alpha$ from $Y^B$. The extension step then depends on the truth values of both $P^X_1(i)$ and $P^Y_1(j)$.

\subparagraph{$P^X_1(i)$ and $P^Y_1(j)$ are both true}
\begin{center}
\begin{tabular}{ c c c | c | c | c}
$P^X_2(i)$ & $P^Y_2(j)$ & $Q$ & $\Delta c$ & $+X^B$ & $+Y^B$ \\
\hline
T & T & & $+2$ & $j,\beta,k_1,k_2$ & $i,\alpha,l_1,l_2$ \\
T & F & & $+1$ & $\beta,k_1,k_2$ & $i,\alpha,l_1$ \\
F & T & & $+1$ & $j,\beta,k_1$ & $\alpha,l_1,l_2$ \\
F & F & $1$ & $+1$ & $j,\beta,k_1$ & $i,\alpha,l_1$ \\
F & F & $0$ & $0$ & $\beta,k_1$ & $\alpha,l_1$
\end{tabular}
\end{center}
where $Q:=F[j,i]+D_X[j,i]F[k_1,i]+D_Y[j,i]F[j,l_1]$.

\subparagraph{$P^X_1(i)$ is true, $P^Y_1(j)$ is false}
\begin{center}
\begin{tabular}{ c c c c | c | c | c}
$P^Y_2(j)$ & $D_Y[j,i]$ & $P^X_2(i)$ & $Q$ & $\Delta c$ & $+X^B$ & $+Y^B$ \\
\hline
T & $1$ & & & $+1$ & $j,\beta,k_1$ & $i,\alpha,l_2$ \\
T & $0$ & T & & $+1$ & $j,k_1,k_2$ & $i,\alpha,l_2$ \\
T & $0$ & F & & $0$ & $j,k_1$ & $\alpha,l_2$ \\
F & $1$ & & & $0$ & $\beta,k_1$ & $i,\alpha$ \\
F & $0$ & T & & $0$ & $k_1,k_2$ & $i,\alpha$ \\
F & $0$ & F & $1$ & $0$ & $j,k_1$ & $i,\alpha$ \\
F & $0$ & F & $0$ & $-1$ & $k_1$ & $\alpha$
\end{tabular}
\end{center}
where $Q:=F[j,i]+D_X[j,i]F[k_1,i]$.

\subparagraph{$P^X_1(i)$ is false, $P^Y_1(j)$ is true}
\begin{center}
\begin{tabular}{ c c c c | c | c | c}
$P^X_2(i)$ & $D_X[j,i]$ & $P^Y_2(j)$ & $Q$ & $\Delta c$ & $+X^B$ & $+Y^B$ \\
\hline
T & $1$ & & & $+1$ & $j,\beta,k_2$ & $i,\alpha,l_1$ \\
T & $0$ & T & & $+1$ & $j,\beta,k_2$ & $i,l_1,l_2$ \\
T & $0$ & F & & $0$ & $\beta,k_2$ & $i,l_1$ \\
F & $1$ & & & $0$ & $j,\beta$ & $\alpha,l_1$ \\
F & $0$ & T & & $0$ & $j,\beta$ & $l_1,l_2$ \\
F & $0$ & F & $1$ & $0$ & $j,\beta$ & $i,l_1$ \\
F & $0$ & F & $0$ & $-1$ & $\beta$ & $l_1$
\end{tabular}
\end{center}
where $Q:=F[j,i]+D_Y[j,i]F[j,l_1]$.

\subparagraph{$P^X_1(i)$ and $P^Y_1(j)$ are both false}
\begin{center}
\begin{tabular}{ c c c c c | c | c | c}
$D_X[j,i]$ & $D_Y[j,i]$ & $P^X_2(i)$ & $P^Y_2(j)$ & $F[j,i]$ & $\Delta c$ & $+X^B$ & $+Y^B$ \\
\hline
$1$ & $1$ & & & & $0$ & $j,\beta$ & $i,\alpha$ \\
$1$ & $0$ & T & & & $0$ & $j,k_2$ & $i,\alpha$ \\
$1$ & $0$ & F & & & $-1$ & $j$ & $\alpha$ \\
$0$ & $1$ & & T & & $0$ & $j,\beta$ & $i,l_2$ \\
$0$ & $1$ & & F & & $-1$ & $\beta$ & $i$ \\
$0$ & $0$ & T & T & & $0$ & $j,k_2$ & $i,l_2$ \\
$0$ & $0$ & T & F & & $-1$ & $k_2$ & $i$ \\
$0$ & $0$ & F & T & & $-1$ & $j$ & $l_2$ \\
$0$ & $0$ & F & F & $1$ & $-1$ & $j$ & $i$ \\
$0$ & $0$ & F & F & $0$ & $-2$ & &
\end{tabular}
\end{center}

\section{Numerical experiments}
\label{sec:experiments}

In the present section, we perform numerical experiments to test and validate our proposed algorithm from \cref{sec:algorithm}.
Specifically, given a graph $G=(V,E)$ and an initial partition size $n$, we run simulated annealing as in \cref{alg:sa} using the update rules for vertex to search for a bipartition with minimal cut rank.
The first two experiments are focused on the general performance of the algorithm, where we consider grid graphs and random graphs, respectively. In the third experiment, we apply the algorithm to \gls{qaoa} as an example for \gls{dqc} with \gls{mbqc}.
In all the result presented here, the temperature schedule $\mathcal{T}$ is chosen linearly in the range from $1.0$ to $0.1$ with a reduction of $0.1$ between each step, unless stated otherwise explicitly.
The code is available online.\footnote{\url{https://github.com/OpenQuantumComputing/min-cutrank}}

\subsection{Gird graphs}

\begin{figure}
\centering
\includegraphics[width=0.45\linewidth]{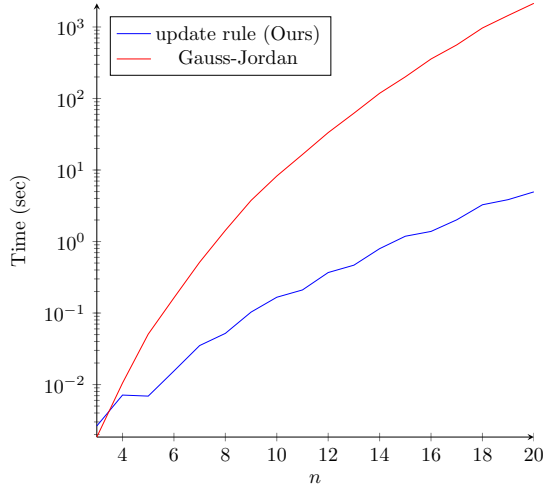}
\caption{Execution times of the simulated annealing algorithm for a balanced bipartition of $n\times n$ grid graphs using naive Gauss-Jordan elimination for calculating the rank in each iteration, vs.~our proposed algorithm using update rules.} \label{fig:sq:time}
\end{figure}

First, we apply our algorithm to $n\times n$ grid graphs. 
Grid graphs are useful to measure how often the annealing algorithm finds a bipartition with minimal cut rank, since the minimum cut rank is known to be $n$ if $n\geq 3$ for balanced cuts. 
For comparison, we evaluate cut ranks both with our proposed update rule as well as with a naive rank calculation using Gauss-Jordan elimination.
The execution times for a balanced bipartition are shown in \cref{fig:sq:time}. 
Clearly, our proposed update rule achieves a better execution time scaling than the naive rank calculation.

\begin{figure}
\centering
\begin{subfigure}[t]{.45\textwidth}
\centering
\includegraphics[width=1.\linewidth]{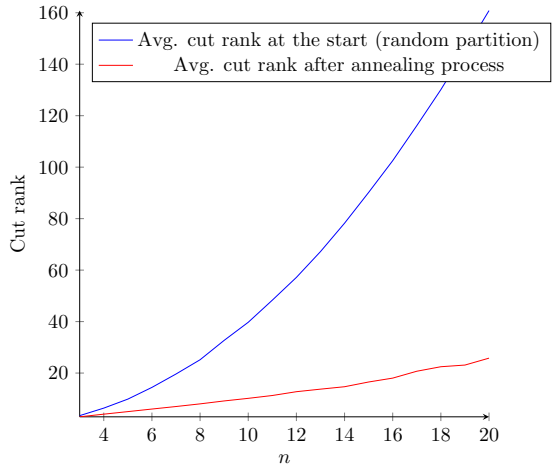}
\caption{Average deviation from known minimal cut rank $n$ after running the annealing algorithm on $100$ grid graphs with random initial partitions.}
\label{fig:sq:deviation:grid1}
\end{subfigure}
\hfill
\begin{subfigure}[t]{.4\textwidth}
\centering
\includegraphics[width=1.\linewidth]{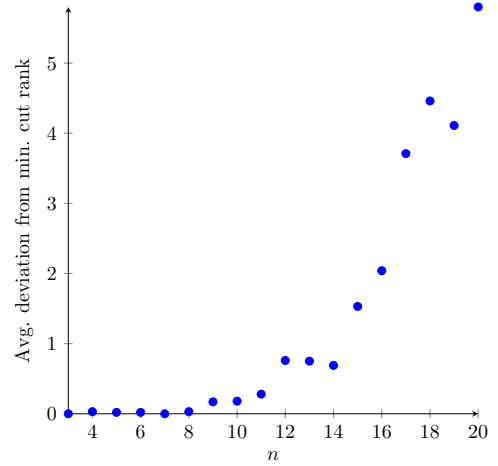}
\caption{Improvement of the cut rank compared to the initial starting partition (smaller is better).}
\label{fig:sq:deviation:grid2}
\end{subfigure}
\caption{Cut rank results from our proposed algorithm using update rules for $n\times n$ grid graphs.} \label{fig:sq:deviation}
\end{figure}

In \cref{fig:sq:deviation:grid1}, we show the improvement of the cut rank from the annealing process using an average over $100$ random initial partitions. 
As can be seen, the simulated annealing helps to decrease the cut rank significantly compared to the starting partition.
In \cref{fig:sq:deviation:grid2}, we compare the cut rank found to the known minimum $n$.
For the largest grids considered here ($20\times 20$) we still find on average partitions that only deviate by approx. 5 from the ideal cut rank.

\subsection{Sparse graphs}

\begin{figure}
\centering
\begin{subfigure}[t]{.44\textwidth}
\centering
\includegraphics[width=1.\linewidth]{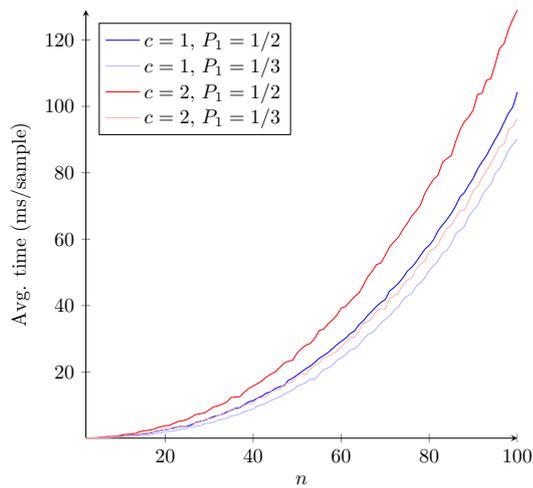}
\caption{average runtimes}
\label{fig:sparse:rank:sparse1}
\end{subfigure}
\hfill
\begin{subfigure}[t]{.44\textwidth}
\centering
\includegraphics[width=1.\linewidth]{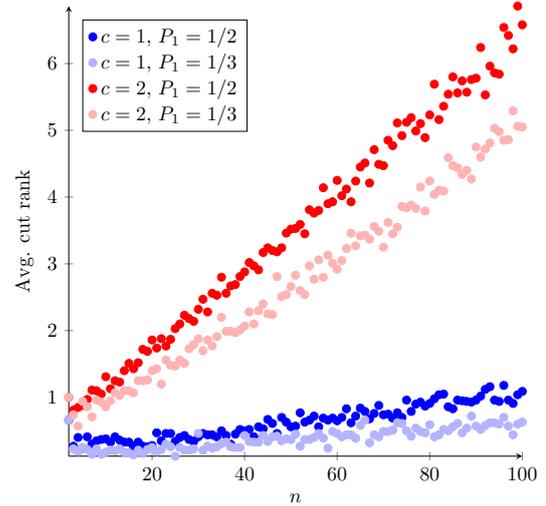}
\caption{average cut rank}
\label{fig:sparse:rank:sparse2}
\end{subfigure}
\hfill
\caption{Results from our proposed algorithm using update rules for $100$ sparse Erd\H{o}s-R\'{e}nyi random graphs $G(n,p)$ for each $n$, where $p=c/n$ and the first partition set has size $P_1n$.} \label{fig:sparse:rank}
\end{figure}

Next, we consider random sparse graphs.
We evaluate the average execution time and final cut rank for the annealing algorithm on $100$ sparse Erd\H{o}s-R\'{e}nyi random graphs $G(n,p)$ for each $n$, where $p=c/n$ and the first partition set has size $P_1n$. 
The average execution times and resulting cut ranks are shown in \cref{fig:sparse:rank}.
From \cref{fig:sparse:rank:sparse1}, it becomes apparent that the scaling behavior of the average execution time is similar for different parameters $c$ and $P_1$.
In \cref{fig:sparse:rank:sparse2}, we observe an overall linear increase of the cut rank with increasing graph sizes.
This is in-line with the known asymptotic scaling of rank-width for random graphs, although technically rank-width only upper-bounds the cut rank for flexible partition sizes with the constraint that the sizes of both sets are between $n/3$ and $2n/3$~\cite{lee2012rank}.
Furthermore, we observe that slightly unbalanced partitions i.e., $(1/3,2/3)$ instead of $(1/2,1/2)$, lead to smaller cut ranks.
This motivates to study more flexible partition schemes in the future.

\subsection{QAOA graphs}

\begin{figure}
    \centering
    \begin{subfigure}{1.\textwidth}
    \centering
    \includegraphics{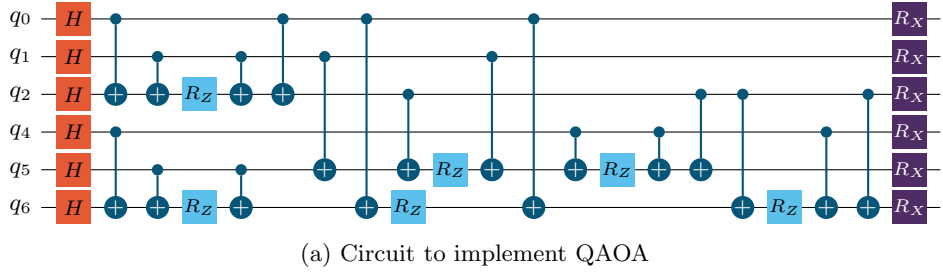}
    \caption{Circuit to implement \gls{qaoa}}
    \label{fig:qaoa_ansatz:circuit}
    \end{subfigure}
    \\[.25cm]
    \begin{subfigure}{1.\textwidth}
    \centering
    \includegraphics[width=0.45\linewidth]{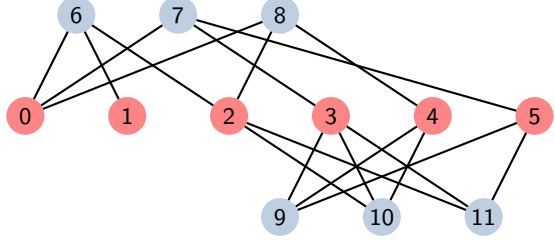}
    \caption{The corresponding \gls{mbqc} computation. The red qubits correpond to the $6$ qubits from the circuit. The blue qubits implement the problem Hamiltonian by measuring in rotated bases.}
    \label{fig:qaoa_ansatz:pattern}
    \end{subfigure}
    \caption{Exemplary circuit and corresponding \gls{mbqc} implementation for \gls{qaoa}.}
    \label{fig:qaoa_ansatz}
\end{figure}

As a prototypical example, we also showcase how our algorithm can be used to distribute an instance of a \gls{qaoa} circuit~\cite{farhi2014}. \Gls{qaoa} has been previously formulated in \gls{mbqc}~\cite{10596340,PhysRevA.106.022437}.
Specifically, we consider a $3$-local Hamiltonian:
\begin{equation} \label{eq:qaoa_ham}
    H := c_1Z_0 Z_1 Z_2+c_2Z_0 Z_3 Z_5+c_3Z_1Z_2Z_4+c_4Z_3Z_4Z_5+c_5Z_2Z_3Z_4+c_6Z_2Z_3Z_5,
\end{equation}
with arbitrary coefficients $c_i \in \mathbb{R}$. We use the standard \gls{qaoa} ansatz with $p=1$:
\begin{equation}
    \ket{\psi} := R_x(\beta_1)\dots R_x(\beta_6) e^{-i\alpha H} \ket{+}^{\otimes 6}.\label{eq:qaoa_ansatz}
\end{equation}
The circuit for this ansatz is shown in \cref{fig:qaoa_ansatz:circuit}.

Next, we derive the \gls{mbqc} protocol that implements the same operation.
For \gls{qaoa}, this is particular simple, since we only need to rewrite terms like $e^{i\theta Z_iZ_jZ_k}$ by measurements on graph states.
This can be achieved by introducing one ancillary qubit for each term in \cref{eq:qaoa_ham} and connecting it with all circuit qubits it has to act on (e.g., with qubit $0,1,2$ for the first term).
These ancillary qubits are then measured in the rotated bases $R_x(\alpha c_i)$, and for outcome $1$, a $Z$-correction has to be applied on all connected circuit qubits. 
This correction can be swapped through the $R_x$ gates in \cref{eq:qaoa_ansatz} resulting in a sign change, whenever the correction has been applied.
The $p=1$ \gls{qaoa} ansatz state, after measurement of the ancillary qubits $q_6$ to $q_{11}$, can be written as
\begin{align}
 \ket{\psi} &= R_x(\beta_1)\cdots R_x(\beta_6) \left(\prod_{i_7 \in N(7)}Z_{i_7}^{s_7}\right) \cdots \left(\prod_{i_{12} \in N(11)}Z_{i_{11}}^{s_{12}}\right) \bra{s_6 \cdots s_{11}} R_x(\alpha_1)\cdots R_x(\alpha_6)  \ket{G}\nonumber\\
 &= R^0_x\left((-1)^{s_6 + s_7}\beta_0\right)R^x_1\left((-1)^{s_6 + s_8}\beta_1\right) R^x_2\left((-1)^{s_6 + s_8 + s_{10} + s_{11}}\beta_2\right)\nonumber\\
 &\quad\times R^x_3\left((-1)^{s_7 + s_9 + s_{10} + s_{11}}\beta_3\right) R^x_4\left((-1)^{s_8 + s_9 + s_{10}}\beta_4\right) R^x_5\left((-1)^{s_7 + s_9 + s_{11}}\beta_5\right) \nonumber\\
  &\quad\times\bra{s_7 \cdots s_{11}} R_x(\alpha_1)\cdots R_x(\alpha_6)  \ket{G},\nonumber
\end{align}
where $\ket{G}$ is the graph state shown in \cref{fig:qaoa_ansatz:pattern} and $s_i \in \{0,1\}$ are the measurement outcomes.

\begin{figure}
    \centering
    \begin{subfigure}[t]{.44\textwidth}
    \centering    
    \includegraphics{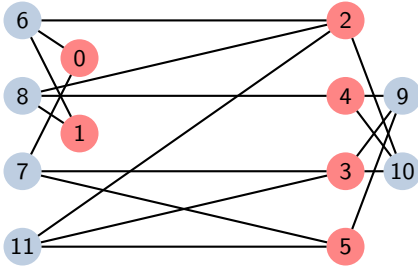}
    \caption{Optimized bipartition with cut rank $3$ from our proposed algorithm.}
    \label{fig:partition:pattern1}    
    \end{subfigure}
    \hfill
    \begin{subfigure}[t]{.44\textwidth}
    \centering    
    \includegraphics{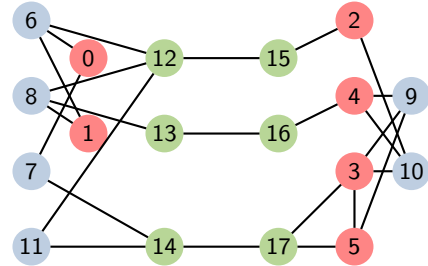}
    \caption{Adding three extra qubits to each partition allows us to split the state.}
    \label{fig:partition:pattern2}    
    \end{subfigure}
    \caption{Preparation of the \gls{qaoa} graph from \cref{fig:qaoa_ansatz:pattern} for \gls{mbqc}-based \gls{dqc} on two \glspl{qpu}.}
    \label{fig:partition}
\end{figure}

In order to distribute $\ket{G}$ across two \glspl{qpu}, we need to select a bipartition $(X,Y)$. Our algorithm finds the partition $X=\{0,1,6,7,8,11\}$ and $Y=\{2,3,4,5,9,10\}$ as shown in \cref{fig:partition:pattern1}, which has cut rank $3$.
As discussed in \cref{sec:distMBQC}, this means that in the \gls{mbqc} procedure, we introduce six ancillary qubits $q_{12},\dotsc q_{16}$, which are pairwise shared on the two \glspl{qpu} as shown in \cref{fig:partition:pattern2}.
Then, measuring the qubits $q_{12},\dotsc,q_{17}$ in the $X$-basis recovers the original graph state up to local unitaries.

\begin{figure}
\centering
\begin{subfigure}[t]{.44\textwidth}
\centering    
\includegraphics[width=1\linewidth]{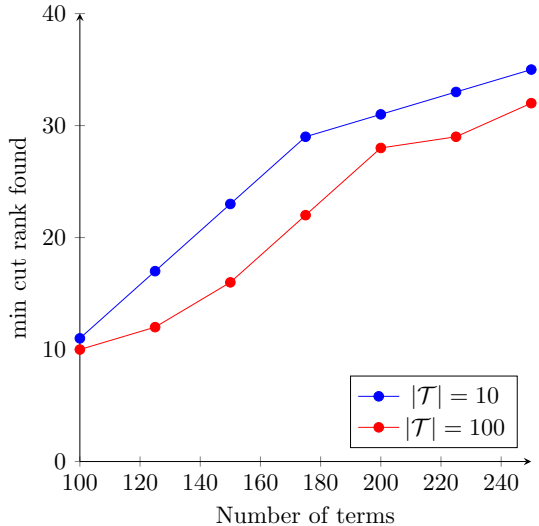}
\caption{Hamiltonian with two-local terms}
\label{fig:qaoa_scaling:weight2}   
\end{subfigure}
\hfill
\begin{subfigure}[t]{.44\textwidth}
\centering    
\includegraphics[width=1\linewidth]{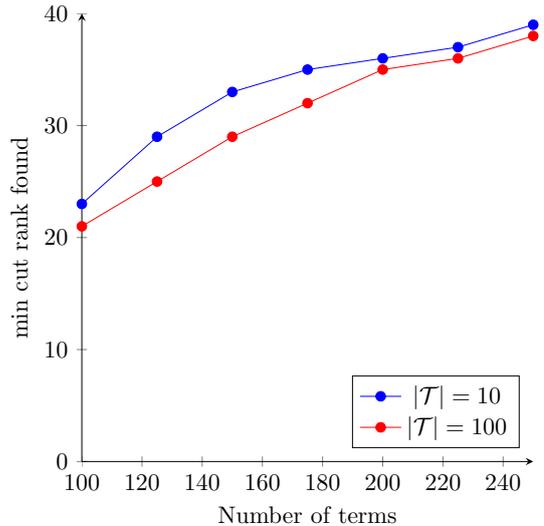}
\caption{Hamiltonian with three-local terms}
\label{fig:qaoa_scaling:weight3}   
\end{subfigure}
\caption{Minimum cut rank found by our algorithm on random \gls{qaoa} instances with $40$ qubits and an increasing number of terms in the Hamiltonian. Two temperature schedules with $|\mathcal{T}| \in \{10,100\}$ steps are considered. The cut rank is upper-bounded by the number of qubits in the respective Hamiltonian ($40$).}
\label{fig:qaoa_scaling}
\end{figure}

To evaluate how our proposed algorithm scales for larger instances, we generate graph states to implement measurement-based \gls{qaoa} as described before using random Hamiltonians with $40$ qubits and a varying number of terms between $100$ and $200$.
We analyze two cases, Hamiltonians with terms of locality $2$ and $3$ (as in \cref{eq:qaoa_ham}).
Furthermore, we also consider two temperature schedules for simulated annealing with ten and 100 steps (still in the range between $0$ and $1$ but with a finer, equidistant spacing), respectively.
The results are shown in \cref{fig:qaoa_scaling}.
Note that the cut rank from partitioning the \gls{mbqc} graph state in this case has an upper bound given by the number of qubits in the circuit model.
As can be seen, our algorithm finds in all cases bipartitions with smaller cut rank than that. 
However, for increasing number of terms, the minimum cut rank found by our algorithm grows. 
This is expected since these cases correspond to circuits with larger depth, meaning more entanglement is shared between the qubits.
As expected, we also observe that running the algorithm with more steps leads to better results.
This suggests that we still do not find the optimal solution here and adding more temperature steps in the simulated annealing algorithm might be necessary, especially for instances with more terms.

\section{Conclusions}
\label{sec:conclusions}

In this work, we introduce an efficient algorithm for the computation of incremental changes in the cut rank of graphs when two nodes are swapped across two partitions.
This can be used in a simulated annealing-based optimization to find fixed-sized bipartitions of a graph with minimum cut rank.
We show how the proposed algorithm can be used to calculate partitions in \gls{mbqc} to distribute a graph state across a network of two \glspl{qpu} minimizing the required number of Bell states shared between the nodes.
This is a first step towards efficient algorithms for graph state partitioning in distributed \gls{mbqc}.
Future work includes extensions to multiple partitions, allowing for partition sizes with defined upper and lower bound sizes (i.e., not fully fixed) and more sophisticated metaheuristics.

\section*{Acknowledgements}
This work was partially supported by the research project \emph{Zentrum f\"ur Angewandtes Quantencomputing} funded by the Hessian
Ministry for Digital Strategy and Innovation and the Hessian Ministry of Higher Education, Research and the Arts (M.H.), and partially by the project
\emph{Quantum Extension of the Norwegian Research Infrastructure (Q-NRI)} funded by the The Research Council of Norway (G.S and K.F.P). 
We thank Sang-il Oum, Tony Huynh, and Franz Fuchs for helpful discussions.
\clearpage

\bibliographystyle{unsrt}
\bibliography{refs} 

\end{document}